\documentclass[final]{ias2}

\usepackage{graphicx} 
\usepackage{multirow}
\usepackage{array} 
\usepackage[arrow, matrix, curve]{xy}
\input{babarsym}
\usepackage{hyperref} 

\begin{document}

\markboth{$B$ decays to baryons}{Torsten Leddig}

\title{$B$ decays to baryons}

\author[HRO]{Torsten Leddig} 
\email{torsten.leddig@uni-rostock.de}
\address[HRO]{Universit\"at Rostock, Institut f\"ur Physik, Universit\"atsplatz 3, 18055 Rostock - Germany}

\begin{abstract}
From inclusive measurements it is known that about $7\%$ of all $B$ mesons decay into final states with baryons.
In these decays, some striking features become visible compared to mesonic decays.
The largest branching fractions come with quite moderate multiplicities of 3-4 hadrons. We note that two-body decays to baryons are suppressed relative to three- and four-body decays.
In most of these analyses, the invariant baryon-antibaryon mass shows an enhancement near the threshold.
We propose a phenomenological interpretation of this quite common feature of hadronization to baryons.
\end{abstract}


 
\maketitle


\section{Decay dynamics}

\begin{figure}[ht]
	\begin{center}
		\includegraphics[height=.27\textwidth]{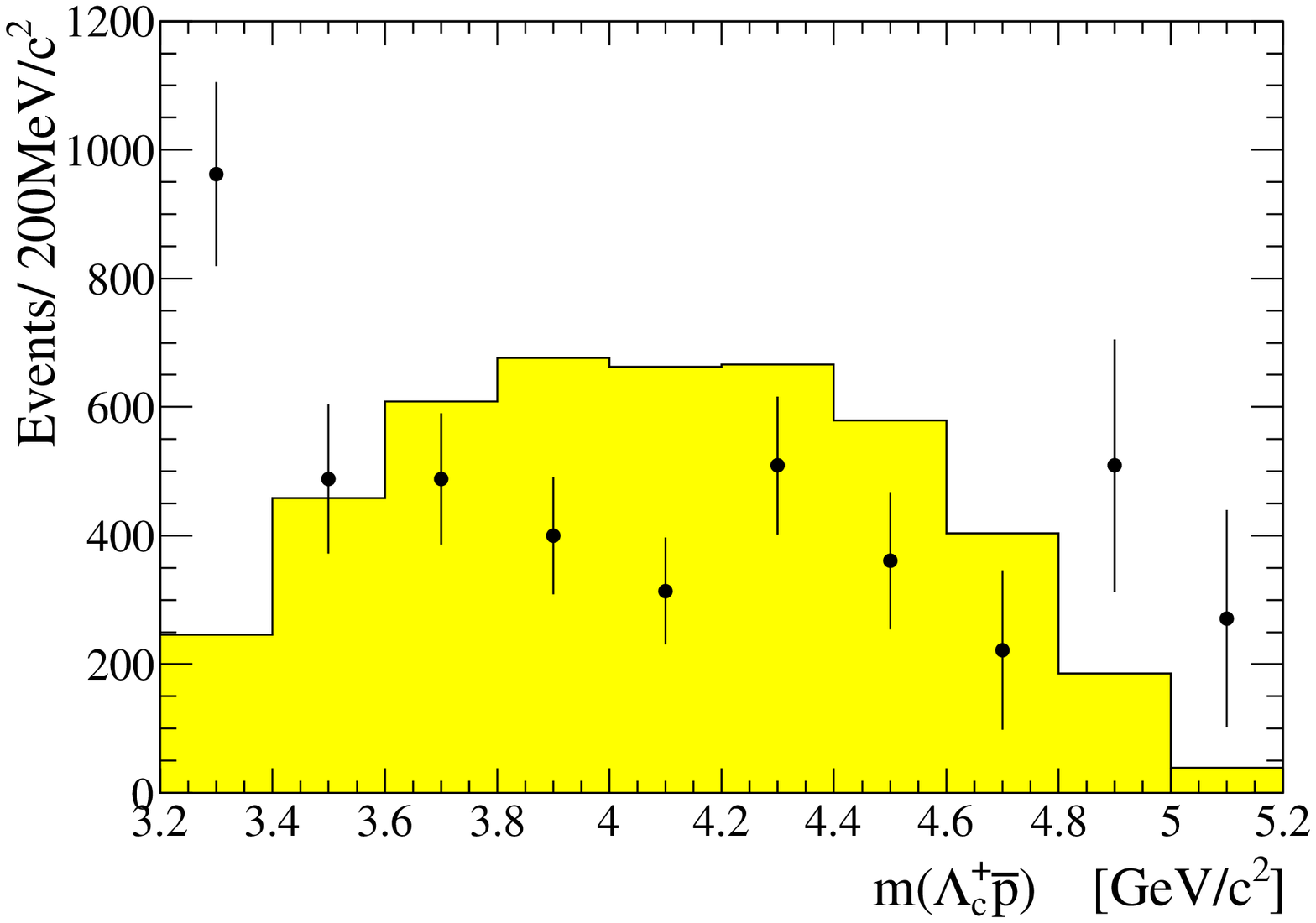}
		\includegraphics[height=.27\textwidth]{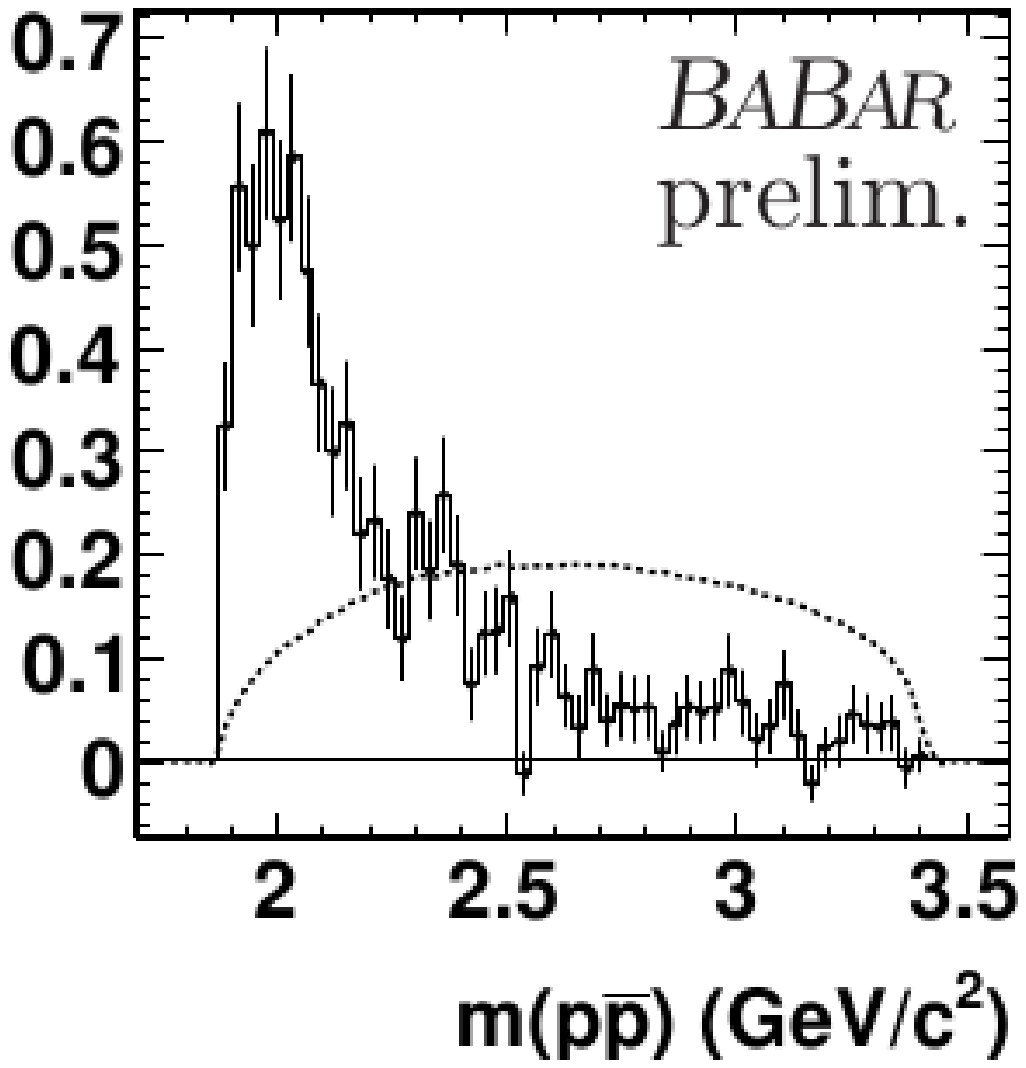}
		\includegraphics[height=.27\textwidth]{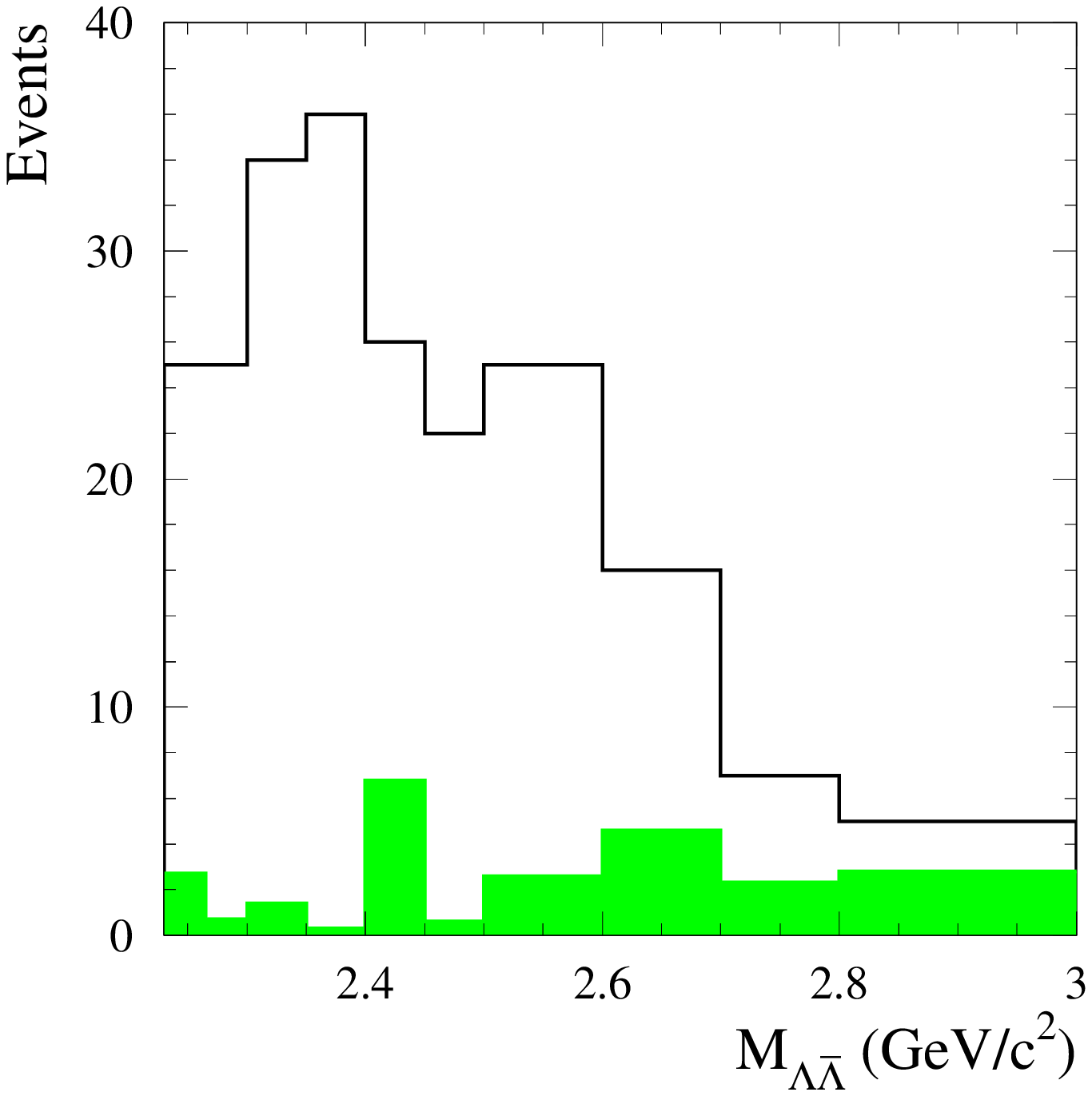}
	\end{center}
	\caption{The enhancement observed at the baryon-antibaryon mass threshold in $\bar{B}^0 \rightarrow \Lambda_c^+ \bar{p} \pi^0$ \cite{Aubert:2010zv} (the yellow histogram represents the phase space expecatation), $\bar{B}^0 \rightarrow D^0 p \bar{p}$ \cite{Aubert:2009qz} (the dotted line represents the phase space expecatation) and $e^+ e^- \rightarrow \Lambda \bar{\Lambda} \gamma$ \cite{Aubert:2007uf} (The shaded histogram shows fitted background).}
	\label{fig:thresh}
\end{figure}

A common feature observed in several $B$ decays to baryons but also outside the $B$-physics sector is an enhancement at the invariant baryon-antibaryon mass threshold which can be seen for three examples in Fig.~\ref{fig:thresh}.

Another feature of these decays is the multiplicity dependence of the branching fractions. Measurements from \babar, Belle and CLEO show that the largest branching fractions for \B decays to baryons come with quite moderate multiplicities. Comparing the branching fractions for $\B \ra \LCp \antiproton \, (n \cdot \pi)$, as shown in Fig. \ref{xy:bf} a rise in the branching fractions up to a multiplicity of five can be observed. The most prominent rise occurs when comparing the two-body mode with the three-body mode for non-resonant decays. For resonant decays ($\Bm\to\SigmaCz(2455)\antiproton$) the difference between the two-body and the three-body mode is not as prominent.
\begin{figure}[ht]
	\begin{center}
		\begin{xy}
			\xymatrix@M=0px@R=10px{
				**[r]\BR\left(\Bzb\to\LCp\antiproton\right)  \ar@/_/[dd]^{{\times 14}} & **[r]\BR\left(\Bm\to\SigmaCz(2455)\antiproton\right) \ar@/_/[dd]^{{\times 4.3}} & **[r]\BR\left(\Bzb\to\LCp\antiproton\right) \ar@/_/[dd]^{{\times 9.5}}\\
				& &	\\
				**[r]\BR\left(\Bm\to\LCp\antiproton\pim\right)_{{\mathsf{non\mbox{-}res}}} \ar@/_/[dd]^{{\times 2.3}} & **[r]\BR\left(\Bzb\to\SigmaCz(2455)\antiproton\pip\right) \ar@/_/[dd]^{{\times 2.9}} & **[r]\BR\left(\Bzb\to\LCp\antiproton\piz\right) \ar@/_/[dd]^{{\times 9.5}}  \\ 
				& &	\\
				**[r]\BR\left(\Bzb\to\LCp\antiproton\pip\pim\right)_{{\mathsf{non\mbox{-}res}}} \ar@/_/[dd]^{{\times 3.4}} & **[r]\BR\left(\Bm\to\SigmaCz(2455)\antiproton\pip\pim\right) & **[r]\BR\left(\Bm\to\LCp\antiproton\pim\piz\right)\\
				& &	\\
				**[r]\BR\left(\Bm\to\LCp\antiproton\pim\pip\pim\right) &   &   \\
			}
		\end{xy}
	\end{center}
	\caption{Relative change of the branching fractions for a subset of baryonic \B decays.}
	\label{xy:bf}
\end{figure}

A comparison of \B decays with a charmed meson in the final state, \linebreak e.g. $\B \ra D^{(*)} \proton \antiproton \, (n\cdot \pi)$\cite{Aubert:2009qz}, shows the highest branching fractions for a multiplicity of four hadrons in the final state. 

\section{Phenomenological interpretation}

Several approaches to explain the suppression of the two-body decay as well as the threshold enhancement in the invariant baryon-antibaryon mass have been suggested. A simple model is given by M. Suzuki \cite{Suzuki:2006nn}. His interpretation is that for a baryon-antibaryon pair in a two-body decay a {\it hard} gluon (highly off mass shell) is needed, while in a decay mode with a higher multiplicity only {\it soft} gluons are needed. In consequence the two-body mode has to be suppressed.
A more detailed model is given by T. Hartmann \cite{hartmann} which can explain the absence of a threshold enhancement in \B decays to baryons. There, all contributing Feynman diagrams are divided into two contributing classes. For convenience for both classes the $W$ exchange can be contracted to an effective four point interaction.

\begin{figure}[ht]
	\begin{center}
		\includegraphics[width=0.45\textwidth]{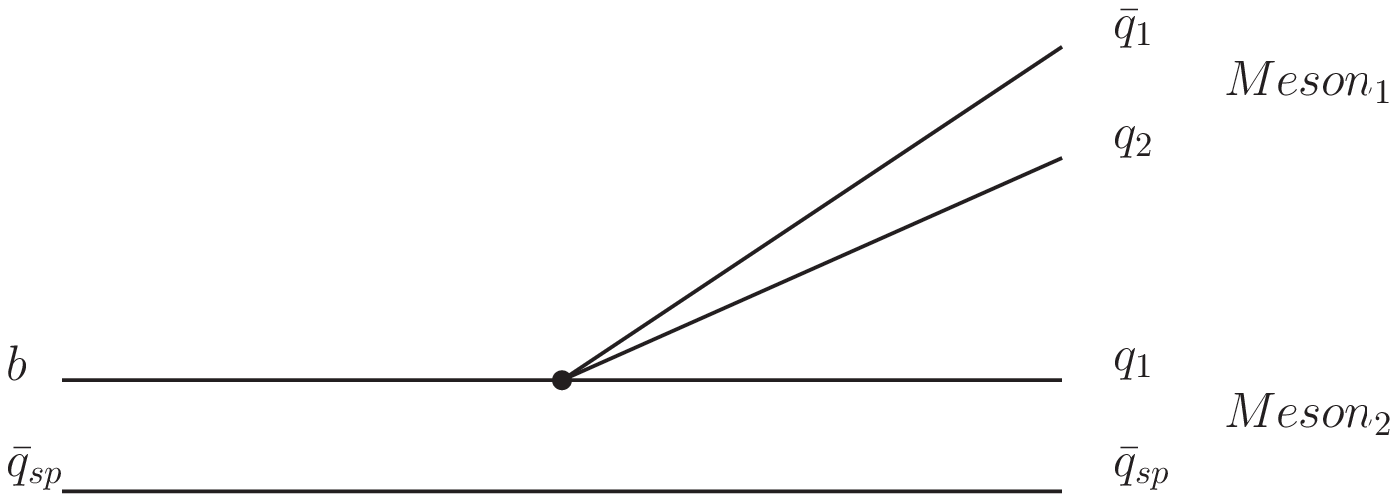}
		\includegraphics[width=0.45\textwidth]{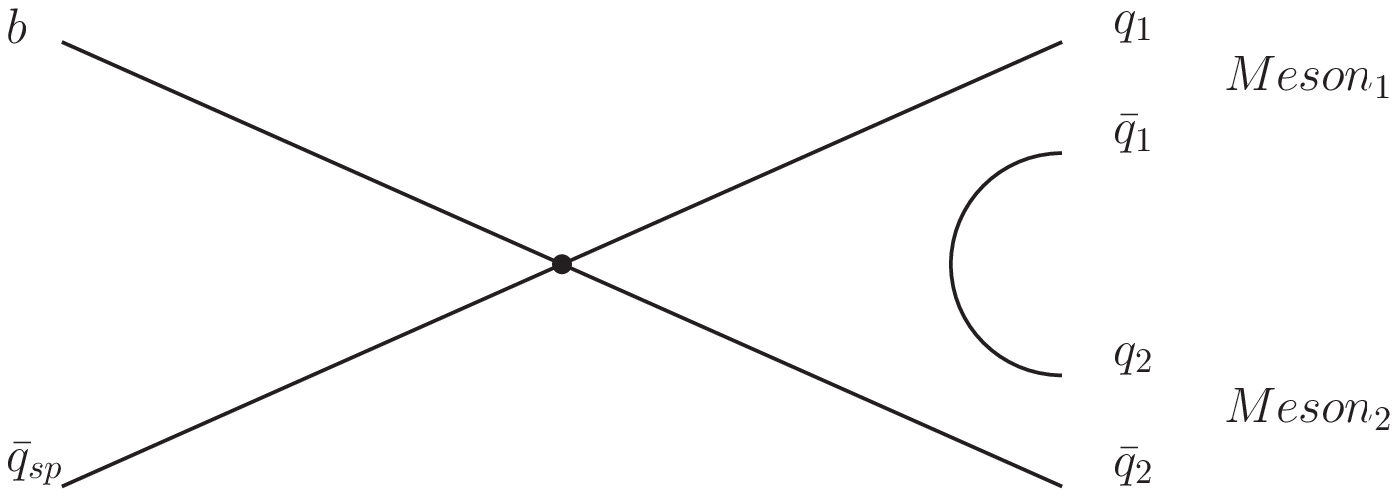} 
	\end{center}
	\caption{Effective Feynman diagrams for the initial meson-meson configuration.}
	\label{fig:Meson}
\end{figure}

In the meson-meson class (Fig. \ref{fig:Meson}) the quarks can be rearranged into a meson-meson configuration with one of the mesons decaying into a baryon-antibaryon pair. In these decays the second meson carries away momentum and reduces the remaining phase space for the baryon-antibaryon pair. This leads to the often observed threshold enhancement. Higher multiplicities are achieved by subsequent decays of the (pseudo-)mesons.

In the second class the quarks are rearranged into a diquark-antidiquark configuration. Since color-confinement requires a quark-antiquark pair created from the gluon field the diquark-antidiquark configuration equals an initial baryon-antibaryon configuration. In consequence no threshold enhancement should be visible for decays proceeding via this type only. An example for a decay proceeding exclusively via this configuration would be $\Bzb \ra \Sigma_c^0 \antiproton \pip$. Possible initial baryon-antibaryon states could be $\Bzb \ra \Sigma_c^0 N$ with $N \ra \antiproton \pip$ or $\Bzb \ra \Lambda_c^{*+} \antiproton$ with $\Lambda_c^{*+} \ra \Sigma_c^0 \pip$.
\begin{figure}[ht]
	\begin{center}
		\includegraphics[width=0.45\textwidth]{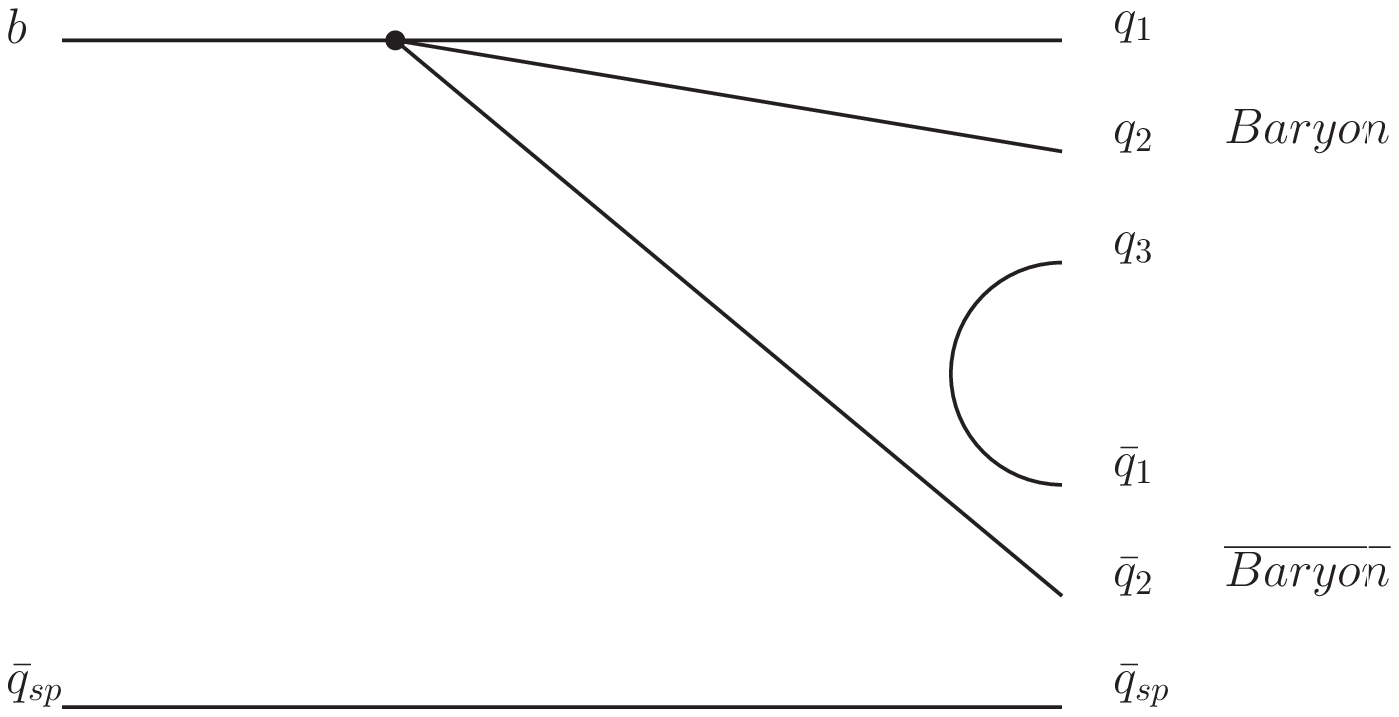} 
		\includegraphics[width=0.45\textwidth]{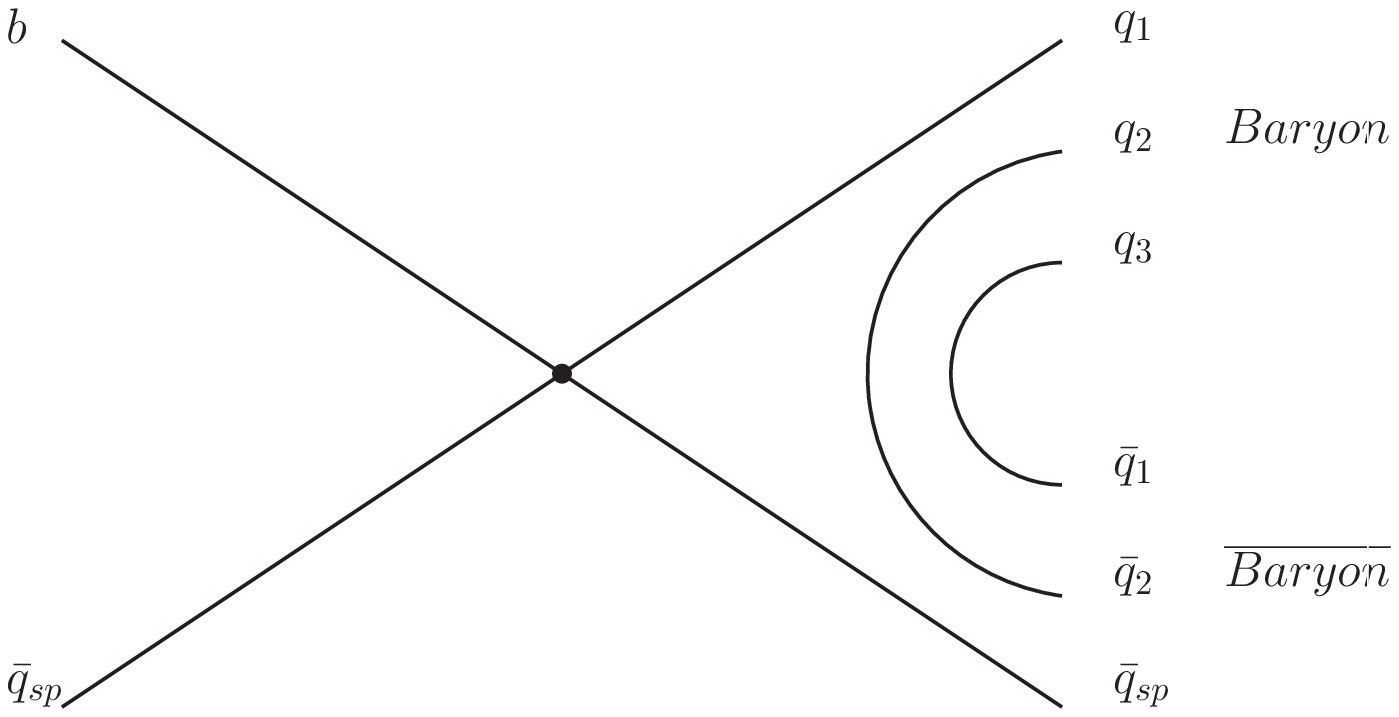} 
	\end{center}
	\caption{Effective Feynman diagrams for the initial diquark-antidiquark configuration.}
	\label{fig:Diquark}
\end{figure} 

\section{Interpreatation of $\Bzb \ra \LCp \Lbar \Km$ results}

\begin{figure}[ht]
	\begin{center}
		\includegraphics[width=0.3\textwidth]{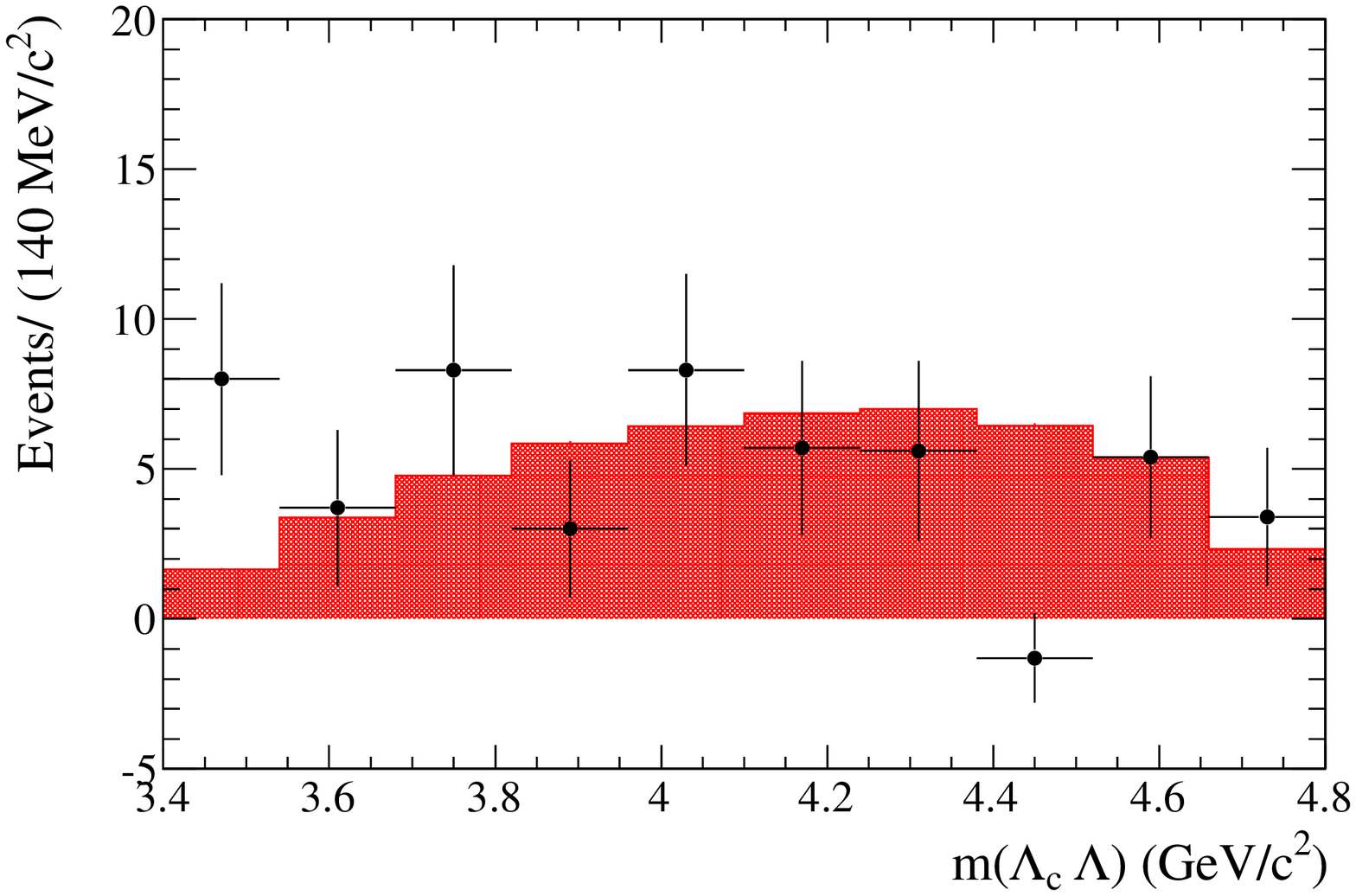} 
		\includegraphics[width=0.3\textwidth]{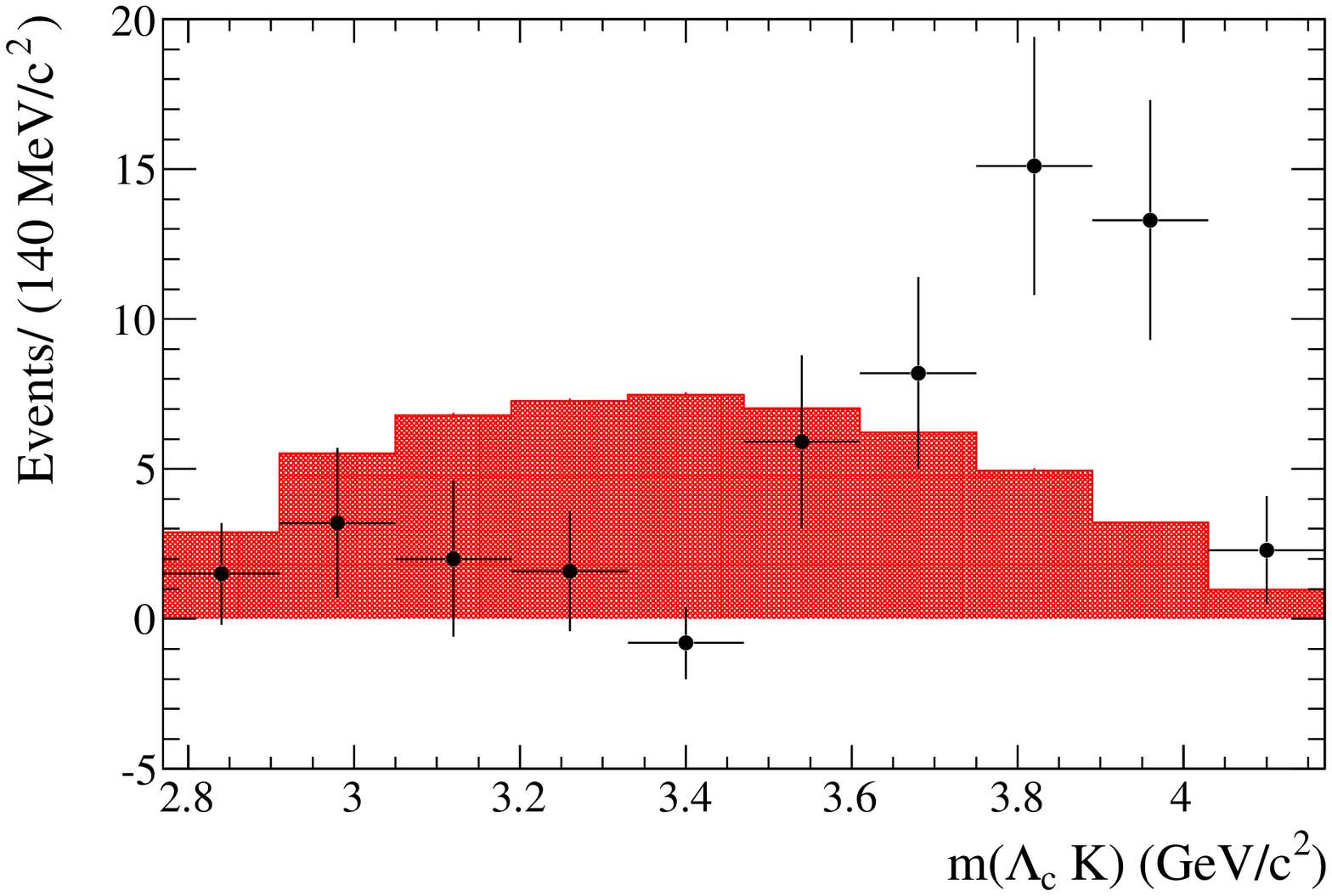}
		\includegraphics[width=0.3\textwidth]{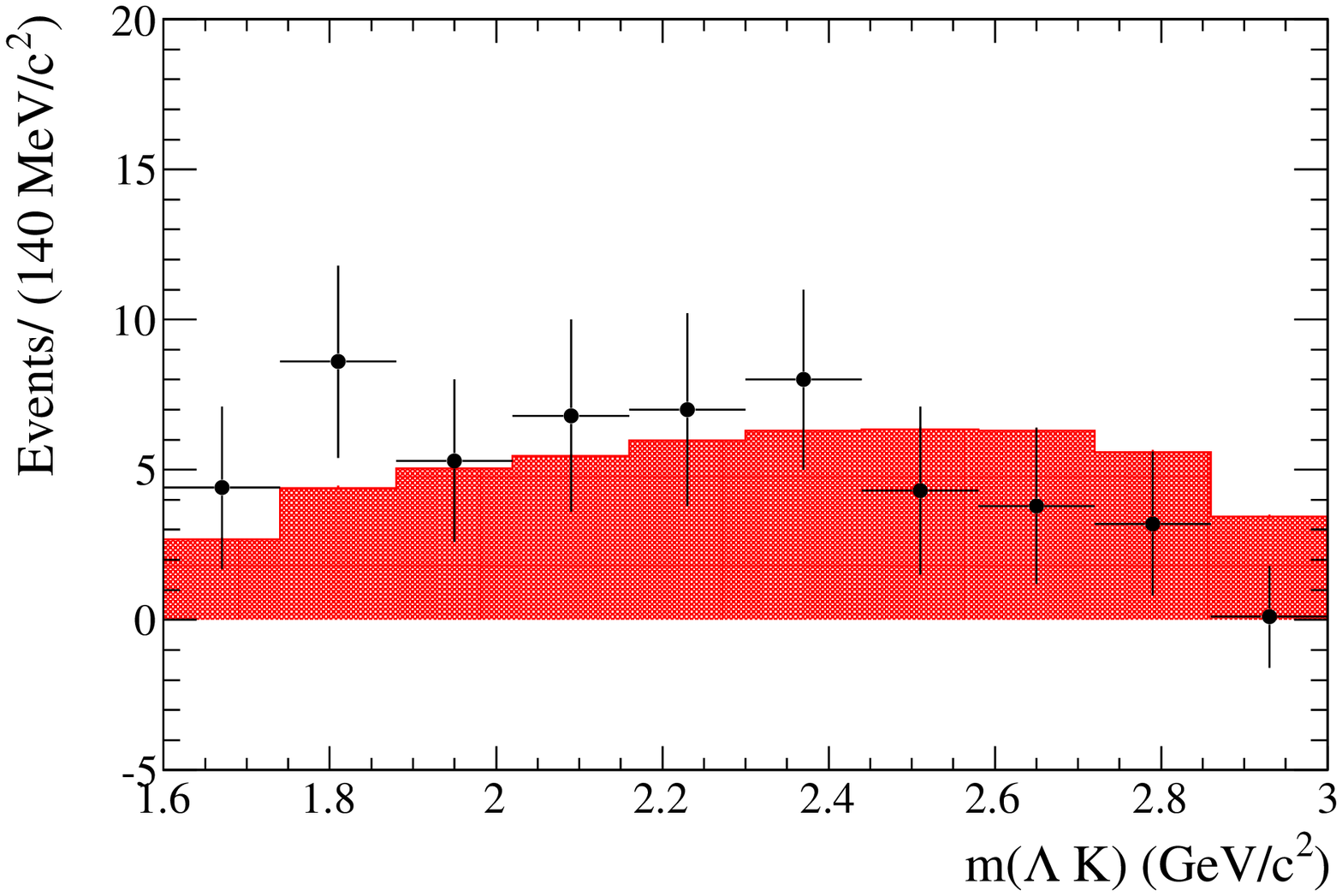} 
	\end{center}
	\caption{Invariant two body mass distributions for the decay $\Bzb \ra \LCp \Lbar \Km$ ($\bullet$) compared the a phase space model (red histogram).}
	\label{fig:LcLK}
\end{figure}

A recent \babar analysis of the decay $\Bzb \ra \LCp \Lbar \Km$ \cite{PhysRevD.84.071102} shows no significant enhancement at the baryon-anibaryon threshold (Fig. \ref{fig:LcLK}). The aforementioned model gives a natural explanation for this. Three Feynman diagrams contribute to this decay. But only one of them can be rearranged into the meson-meson configuration which is necessary for the threshold enhancement. Depending on the relative strengths of the three contributing Feynman diagrams this provides a natural explanation for the absence of a strong enhancement.

\bibliographystyle{pramana}
\bibliography{references}

\end{document}